\documentclass[12pt]{article}
\usepackage{amsmath,amssymb,physics}
\usepackage[mathscr]{eucal}
\usepackage{MnSymbol}
\usepackage{mathtools}

\usepackage{color}  % Red, Blue, Yellow, Green
  \definecolor{R}{rgb}{1.0,0,0}
  \definecolor{B}{rgb}{0.1,0.1,0.9}
  \definecolor{Y}{rgb}{0.8,0.5,0.2}
  \definecolor{G}{rgb}{0,0.79,0}
\usepackage[normalem]{ulem}
\newcommand{\added}[2][R]{{\color{#1}#2}}

\def\bb{\mathbb}

\def\R{{\bb R}}                             % Set of real numbers
\def\nrN{N}                                                       %  number of objects
\def\cf{\mathfrak{n}}                                         %  counting function
\def\cfu{\cf_\star}                                              %  universal counting function
\def\efN{\mathscr{N}}                                        %  effective number of objects
\def\efNm{\efN_\star}                                        %  minimal effective number of objects
\def\mN{\mathscr{N}_+}                                    %  mathematical number
\def\mn{\mathrm{n}_+}                                      % mathematical density
\def\w{c}                                                              %  counting weight
\def\W{C}                                                             % counting vector
\def\setW{\mathcal{C}}                                        % set of counting vectors 
\def\Nmaps{{\mathfrak N}}                                   % set of N-maps
\def\opO{\hat{O}}                                                  % operator O
\def\ketpsi{{\mid \! \psi \,\rangle}}                          % ket-psi
\def\keti{{\mid \! i \,\rangle}}                                   % i-psi
\def\sdm{{\Omega}}                                                % spectral domain
\def\efVs{{\mathscr{V}}}                                          % effective spectral volume

\begin{document}

\begin{center}

{\Large{\bf A Different Angle on Quantum Uncertainty}} \\
\vspace*{.12in}
{\Large{\bf (Measure Angle)}} \\
\vspace*{.24in}
{\large{Ivan Horv\'ath$^{1,a}$ and Robert Mendris$^{2,b}$}}\\
\vspace*{.24in}
$^1$University of Kentucky, Lexington, KY, USA\\
\vspace*{.04in}
$^2$Shawnee State University, Portsmouth, OH, USA

\vspace*{.25in}
{\large{June 18  2019}}

\end{center}

\vspace*{0.20in}

\begin{abstract}

\bigskip
\noindent
The uncertainty associated with probing the quantum state is expressed as 
the effective abundance (measure) of possibilities for its collapse. New kinds of uncertainty 
limits entailed by quantum description of the physical system arise in this manner.


\renewcommand{\thefootnote}{}
\footnotetext{\hspace*{-.35cm} Talk at the 7th International Conference on New Frontiers 
in Physics (ICNFP2018), 4-12 July 2018, Kolymbari, Crete. 
To appear in {\em Proceedings} (MDPI).\\ 
\indent ${}^a${\tt ihorv2@g.uky.edu}, $\,{}^b${\tt rmendris@shawnee.edu}
}
\end{abstract}

\vfill\eject

\section{Introduction}

One could easily imagine the results of this presentation being reported at a physics conference long ago, 
well before Alice, Bob and Charlie were part of quantum discourse. In fact, it would have been natural for 
this to happen when the Copenhagen interpretation of quantum mechanics (QM) was only 
emerging~\cite{Hei27A}. One reason it did not occur may be that the needed association of probability 
and measure was not developed or appreciated at the time, although it very well could have been. Here 
we obviously do not mean the use of measure theory to formalize probability \cite{Kol33A}. 
Rather, what we have in mind is a generalization of measure by means of probability: the extension of 
the measure map $\mu=\mu(A)$ onto a larger domain $\mu=\mu(A, \pi)$, where $\pi$ is a probability 
measure on $A$. The role of $\pi$ is to specify the ``relevance'' for various parts (measurable subsets) 
of $A$. The desired extension is then chiefly driven by two requirements. 
(i) $\mu(A, \pi)$ should decrease relative to $\mu(A)$ in response to $\pi$ favoring 
certain parts of $A$, so that $\pi$ involving more concentrated probability entails 
larger reduction (monotonicity with respect to cumulation).
(ii) $\mu(A, \pi)$ should remain strictly measure-like in that the original additivity 
relation involving sets $A$ and $B$ generalizes into one involving pairs $(A,\pi_A)$ 
and $(B,\pi_B)$, for all $\pi_A$ and $\pi_B$. 
If (i) and (ii) can be accommodated simultaneously and together with few other basic
requirements, then $\mu(A, \pi)$ defines a meaningful {\em effective measure} of 
$A$ with respect to $\pi$. Such quantifier could then be used in correspondingly 
wider contexts, but with essentially the same meaning and significance, as that of 
an ordinary measure. Being a framework for assigning probabilities to events, 
quantum mechanics would be among prime natural settings for its use.

Whether the outlined general approach materializes into fruitful enrichment of 
QM depends on the existence and multitude of the above extensions for relevant measures.
In Ref.~\cite{Hor18A}, released during the ICNFP 2018 meeting, the extension program 
was completely carried out for the foundational case of counting measure. It is 
the surprising results of this analysis that suggest, among other things, a qualitatively 
new outlook on quantum uncertainty \cite{Hei27A,Ken27A} that we point out 
in this presentation. Since the ICNFP 2018 meeting, the ensuing concept of 
{\em measure uncertainty} ($\mu$-uncertainty) has been fully developed in 
Ref.~\cite{Hor18B}. Over the course of that process, effective measures 
were also defined for subsets of $D$-dimensional Euclidean space $\R^D$ 
with Jordan content, i.e. whose ``volume" is expressible as Riemann integral.

Since the discrete case involves a somewhat specific language, we summarize
the correspondence with the general case before we start. A ``measurable set $A$'' 
becomes simply a ``collection of $\nrN$ objects'', and $\mu(A)$ corresponds to $\nrN$. 
A probability measure $\pi$ is specified by the probability vector 
$P=(p_1,p_2,\ldots,p_\nrN)$. The construction of effective measure extension 
$\mu=\mu(A,\pi)$ then turns into the construction of function $\efN=\efN[P]$ 
interpreted as the effective total (effective count). The theory of these objects 
is referred to as the {\em effective number theory} since $\efN$ retains certain 
algebraic features of integers~\cite{Hor18A}. In the same vein, functions $\efN$ 
are called the effective number functions (ENFs).

In the first part of the presentation, we will describe the key features and 
results of the effective number theory, which is a crucial stepping stone for our 
arguments regarding quantum uncertainty. One natural approach to constructing 
the theory is to view it as a tool to solve a generic counting problem of quantum 
mechanics, which we refer to as the quantum identity problem~\cite{Hor18A}. 
In particular, consider the state $\ketpsi$ and the basis 
$\{\, \keti \} \equiv \{ \, \keti  \mid i=1,2,\ldots,\nrN \,\}$ in $\nrN$-dimensional 
Hilbert space. Is it meaningful to ask how many basis states 
$\keti$ (``identities'' from $\{\, \keti \}$) are effectively contained in $\ketpsi$? 
This question can be phrased in many equivalent ways that directly relate to 
a particular application of interest. For example, in the context of describing 
localization, the inquiry would be about how many states from 
$\{\, \keti \}$ is $\ketpsi$ effectively spread over. On the other 
hand, in assessing the efficiency of a variational calculation involving eigenstate 
$\ketpsi$ and basis $\{\, \keti \}$, we would be concerned with how many $\keti$ 
effectively describe $\ketpsi$. Yet, it is the same quantum identity question arising 
in all these situations, and it entails seeking the maps 
\begin{equation}
    \ketpsi   \; , \;  \{ \, \keti \}  \qquad \longrightarrow \qquad 
    \efN[\, \ketpsi, \{ \,\keti \} ] 
    \label{eq:010}        
\end{equation}
consistently assigning the effective totals. Hence, in the context of the quantum 
identity problem, the ''objects" in the discrete sets are the basis states $\keti$ and 
their probabilistic weight is encoded in $\ketpsi$ by the quantum-mechanical rule 
$p_i = \,\mid \!\! \langle \,i \!\mid \! \psi \,\rangle \!\! \mid^2$. 
The effective number function $\efN$ is thus a map whose domain consists of probability 
vectors $P$ so constructed.\footnote{We note here that, as a part of further research, 
we suggest to look at many other active areas of physics to find applications of 
the quantum identity problem. Given the nature of their focus, the appropriately 
formulated questions may arise in the context of Anderson localization, quantum 
chaos, quantum computing/information, entanglement, thermalization, topological 
insulators, holography and possibly others.}
In the second part of this presentation, we argue that effective number theory 
leads to a novel understanding of quantum uncertainty. In fact, the association of 
effective measure and uncertainty constitutes a conceptual thread connecting all 
of the mentioned quantum applications.   

%%%%%%%%%%%%%%%%%%%%%%%%%%%%%%%%%%%%%%%%%%
\section{Effective Number Theory}

\begin{figure}[t]
\begin{center}
    \centerline{
    \hskip 0.00in
    \includegraphics[width=15.0truecm,angle=0]{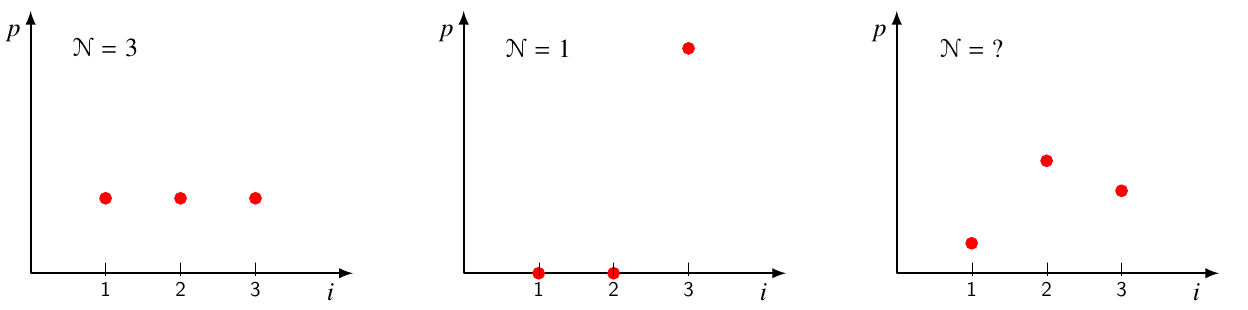}
     }
     \vskip -0.0in
     \caption{The effective number assignments for uniform (left) and the $\delta$-function (middle) probability distributions 
     are postulated. Is there a well-founded prescription(s) for generic distributions (right)?}
     \label{fig:1}   
     \vskip -0.10in
\end{center}
\end{figure}

Although we are concerned with describing an abstract theory, it is useful to keep a concrete 
physical situation in mind when doing so. 
We will use the elementary example of \added[G]{a} spinless lattice Schr\"odinger particle, 
described by wave function 
$\ketpsi \rightarrow ( \psi(x_1),\ldots,\psi(x_\nrN) )$, for this purpose. 
Thus, $\nrN$ is the number of lattice sites and $p_i \!=\! \psi^\star \psi(x_i)$ refers to 
the probability of detecting the particle at position $x_i$. In the language of the 
quantum identity problem \eqref{eq:010}, we ask how many position basis states 
$\keti$ is $\ketpsi$ effectively composed of? The situation is exemplified in Fig.~\ref{fig:1}, 
showing three probability distributions on the lattice with three sites. 
Clearly, all ENFs should assign $\efN \!=\!3$ to uniform distribution (left panel) and 
$\efN \!=\!1$ to $\delta$-function distribution (middle panel). 
The problem in question boils down to establishing whether 
well-founded effective numbers can be assigned to generic distributions, 
such as the one shown in the right panel\added[G]{.}

Our approach to this issue is to develop an axiomatic definition of the set $\Nmaps$ containing all effective 
number functions $\efN$, and then analyze its properties \cite{Hor18A}. Two of the axioms play a prominent 
role in shaping the possible ENFs, namely additivity and monotonicity with respect to ``cumulation". 
The latter is closely related to Schur concavity. Since both need some care in their formulation, we will discuss 
them in somewhat more detail than other requirements. It should be noted that additivity, while crucial for 
a measure-like concept, was not part of related considerations in the past.

\subsection{Additivity}

Consider the lattice Schr\"odinger particle on the lattice with $\nrN$ sites, in a state producing the probability 
vector $P=(p_1,\ldots,p_\nrN)$. Similarly, let the particle be restricted to a non-overlapping lattice of $M$ 
sites, generating the probabilities $Q=(q_1, \ldots, q_M)$. Then there exists a state of the particle on 
the combined lattice (see Fig.~2) that leads to the probability distribution
\begin{equation}
      P \diamondplus Q  \;\;\,\equiv\;\;\,  \frac{N}{N+M} \,P \,\boxplus\,  \frac{M}{N+M} \, Q  
      \;\;\,=\;\;\, \frac{N}{N+M} \, (p_1,\ldots,p_\nrN) \;\boxplus\; \frac{M}{N+M} \, (q_1,\ldots , q_M)   \quad
      \label{eq:020}              
\end{equation}
on the position basis of the combined system. Here $\boxplus$ denotes the concatenation operation, namely 
$(a_1,\ldots,a_\nrN) \boxplus (b_1,\ldots, b_M) \equiv (a_1,\ldots,a_\nrN,b_1,\ldots,b_M)$. 
Note that $P \diamondplus Q$ doesn't change the weight ratios for position states inside the two parts of 
the system, thus preserving the individual distribution shapes. It also properly corrects weight ratios of position 
pairs from distinct parts by their respective ``measures" (lattice sizes $\nrN$ and $M$). While the nominal size 
of the combined system is trivially given by $N+M$ (ordinary count), its effective size in a said state 
(effective count) also has to be additive, yielding the corresponding condition for ENFs, namely
\begin{equation} 
      \efN[P \diamondplus Q] \;\equiv\; \efN[P] \,+\, \efN[Q]
      \label{eq:030}                     
\end{equation}
The above equation has to be satisfied for all $P$ and all $Q$.

\begin{figure}[t]
\begin{center}
    \centerline{
    \hskip 0.00in
    \includegraphics[width=8.0truecm,angle=0]{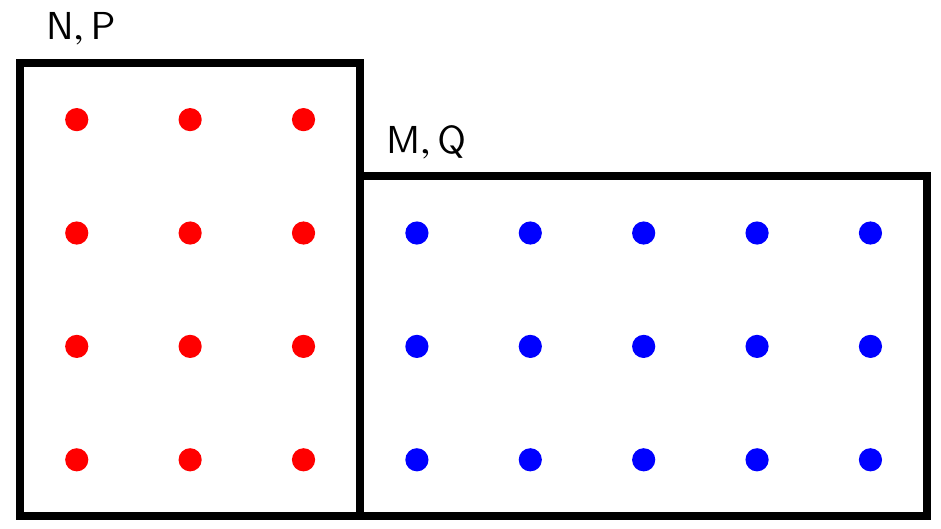}
     }
     \vskip -0.0in
     \caption{Composing the probability distributions generated by lattice Schr\"odinger particle.  Parts 
     characterized by $\nrN$, $P$ (red) and $M$, $Q$ (blue) combine into a total described by 
     $\nrN + M$,  $P \diamondplus Q$. See the discussion in the text.}
     \label{fig:2}   
     \vskip -0.10in
\end{center}
\end{figure}

The measure conversion factors, like those appearing in Eq.~\eqref{eq:020}, can be eliminated 
from all relevant expressions by simply working with {\em counting vectors} $\W$ rather than 
probability vectors $P$, namely
\begin{equation}
    P = (p_1, \ldots, p_\nrN)   \quad \longrightarrow \quad \W \,\equiv \nrN P 
    \,=\, (\w_1, \ldots , \w_\nrN) \,=\,  (\nrN p_1, \ldots , \nrN p_\nrN)
    \label{eq:040}          
\end{equation} 
Formally, the set of counting vectors $\setW$ is defined as
\begin{equation}
    \setW \,=\, \bigcup_\nrN \,\setW_\nrN   \qquad , \qquad 
    \setW_\nrN  \,=\,  \bigl\{ \, (\w_1,\w_2, \ldots, \w_\nrN) \;\mid\;  \w_i \ge 0 \,,\,\, 
    \sum_{i=1}^\nrN \w_i = \nrN \, \bigr\}  \quad
    \label{eq:050}
\end{equation}
Clearly, if $\W \in \setW_\nrN$ and  $B \in \setW_M$, then the counting vector associated with 
the combined system is simply $\W \boxplus B \in \setW_{\nrN+M}$, which is to be compared 
with Eq.~\eqref{eq:020}. Thus, from now on, we treat ENFs as maps whose domain is $\setW$, 
namely $\efN=\efN[\W]$ , $\W \in \setW$. The additivity condition then reads~\cite{Hor18A}
\begin{equation}
     \efN \bigl[ \W \boxplus B \bigr]  \;=\; 
     \efN \bigl[ \W \bigr]  \,+\, \efN \bigl[ B \bigr]   
     \qquad , \qquad \forall \,\W , B \in \setW
     \tag{A}
     \label{eq:add}         
\end{equation}

\subsection{Monotonicity}

The purpose of monotonicity property is to ensure that, given a pair of distributions 
$\W$, $B \in \setW_\nrN$, the one with more cumulated weights won't be assigned 
a larger effective number. To formulate the requirement, it is important to realize that 
all $\W$, $B$ cannot be readily compared by degree of their cumulation. In the left 
panel of Fig.~3 we show an example of a comparable pair. For this purpose, 
weight entries were ordered in a decreasing manner so that the cumulation 
center is on the left. Distribution $\W$ is more cumulated since it can be obtained from 
$B$ by transfer of weight (flow) directed toward the cumulation center at all times. 
On the other hand, the distributions shown in the right panel cannot be compared without 
introducing ad hoc assumptions, since flow in both directions is needed to perform 
the needed deformation. Thus, the universal notion of monotonicity is only concerned 
with properly treating the situation on the left.

\begin{figure}[t]
\begin{center}
    \centerline{
    \hskip 0.00in
    \includegraphics[width=11.0truecm,angle=0]{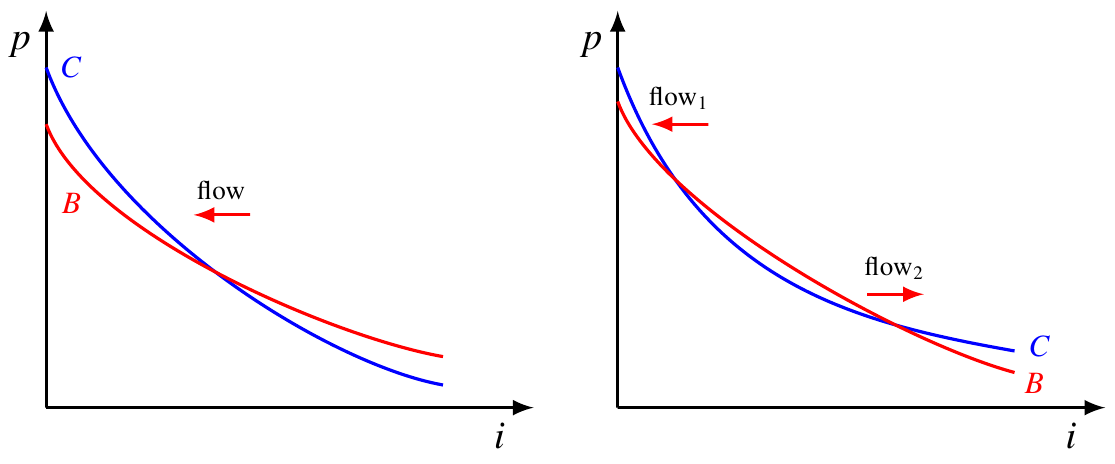}
     }
     \vskip -0.0in
     \caption{In the left panel, distribution $C$ is more cumulated than distribution $B$. Distributions 
     in the right panel cannot be readily compared by their cumulation. Note that discrete dependencies 
     were replaced with continuous ones for better clarity.}
     \label{fig:3}   
     \vskip -0.27in
\end{center}
\end{figure} 

It is straightforward to check that, for any pair of such comparable discrete distributions, the deformation in question can be carried out as a finite sequence of pair-wise weight exchanges, each transferring weight toward the center of cumulation. 
Since performing such an elementary operation on $\W$ produces $\W'$ such that ($\W$, $\W'$) is a comparable pair, the monotonicity requirement is entirely captured by the set of conditions concerning these elementary operations, namely
\begin{equation}
      \efN(\ldots \w_i - \varepsilon  \ldots  \w_j + \varepsilon  \ldots)   \;\le\; 
      \efN(\ldots  \w_i  \ldots  \w_j  \ldots)   
      \quad\; , \quad\;  \forall \,\w_i \le \w_j    \quad\; , \quad\;  \forall \, \varepsilon \,:\, 0 \le \varepsilon \le \w_i
      \quad   \tag{M$^-$}
      \label{eq:mon-} 
\end{equation}                        
\eqref{eq:mon-} monotonicity is designed to identify functions respecting cumulation. To place its meaning in a more 
conventional context, we point out that imposing it in conjunction with symmetry \eqref{eq:sym} (see below) results 
in a well-known property of Schur concavity \cite{Arn87A}. 

\subsection{Effective Number Functions}

Apart from additivity \eqref{eq:add} and monotonicity \eqref{eq:mon-}, the axioms defining effective number 
functions $\efN = \efN[\W]$ incorporate intuitive and easily formulated features such as symmetry, continuity, 
and previously mentioned ``boundary values'' associated with uniform and $\delta$-function distributions. 
The complete list of these additional requirements is formally specified below.
\begin{equation}
    \efN(\ldots \w_i  \ldots \w_j  \ldots)  \,=\,  
    \efN(\ldots \w_j  \ldots  \w_i  \ldots)    
    \qquad , \qquad  \forall\, i \ne j  \tag{S}
     \label{eq:sym}             
\end{equation}    
\begin{equation}
    \efN = \efN[\W]  \;\, \text{is continuous on} \;\, \setW_\nrN  \quad , \quad \forall \,\nrN \tag{C}
     \label{eq:con}             
\end{equation}    
\begin{equation}
     \efN(1,1,\ldots,1) = \nrN      
     \qquad , \qquad  (1,1,\ldots,1) \in \setW_\nrN  
     \quad,\quad  \forall \, \nrN    \tag{B1}
     \label{eq:bou1}                   
\end{equation}
\begin{equation}
     \efN(\nrN,0,\ldots,0) = 1        
      \qquad , \qquad  (\nrN,0,\ldots,0) \in \setW_\nrN   
      \quad,\quad  \forall \, \nrN       \tag{B2}
     \label{eq:bou2}                   
\end{equation}
\begin{equation}
     1 \, \le \, \efN[\W] \, \le \, \nrN   \qquad , \qquad \forall \, \W \in \setW_\nrN 
      \quad , \quad   \forall \, \nrN       \tag{B} 
     \label{eq:bou}                                                                 
\end{equation}
It can be shown that \eqref{eq:bou1} and \eqref{eq:bou} follow from the remaining five requirements, leading 
to \added[G]{a} definition of ENF set $\Nmaps$ as a collection of real-valued functions $\efN = \efN[\W]$ on 
$\setW$, satisfying \eqref{eq:add}, \eqref{eq:mon-}, \eqref{eq:sym}, \eqref{eq:bou2} and \eqref{eq:con}.

Each ENF contained in $\Nmaps$ provides a consistent scheme to assign effective totals 
(effective counting measures) to sets of objects endowed with counting/probability weights. 
It should be pointed out in that regard that none of the quantifiers currently used as substitutes for effective 
numbers, such as participation number \cite{Bel70A}, exponentiated Shannon entropy \cite{Sha48A} 
or exponentiated R\'enyi entropies~\cite{Ren60A}, is \eqref{eq:add}-additive. However, they do respect 
all the other axioms defining $\Nmaps$. 

\subsection{The Minimal Amount}

Interestingly, the effective number theory based on the above definition of $\Nmaps$ can be entirely 
solved~\cite{Hor18A}. In fact, all ENFs were explicitly found, and the structural properties of $\Nmaps$ 
were established. These results are summarized by Theorems 1,2 of Ref.~\cite{Hor18A}. To convey the aspects 
needed here, we first define the function $\mN$ on $\setW$, counting the number of objects with non-zero 
weights, namely 
\begin{equation}
     \mN[\W] = \sum_{i=1}^\nrN \mn(\w_i)    \qquad\qquad
     \mn(\w)  \,=\,
     \begin{cases} 
      \;0 \;,   &   \;  \w=0 \\[8pt]
      \;1 \;,   &   \;  \w > 0
  \end{cases} 
  \label{eq:060}         
\end{equation}
Note that $\mN$ is not an ENF due to the lack of continuity. 
In addition, the following function $\efNm$ on $\setW$ is important 
in the context of effective number theory and the Theorem below.
\begin{equation}
    \efNm[\W]  \,=\, \sum_{i=1}^\nrN \cfu(\w_i) 
    \qquad\qquad   \cfu(\w)  \, = \,   \min\, \{ \w, 1 \}  
    \label{eq:065}             
\end{equation}

\bigskip

\noindent {\bf Theorem.}
{\em There are infinitely many elements in $\Nmaps$ including $\efNm$. 
Moreover, for every fixed $\W \in \,\setW$} 
\smallskip
\begin{equation}     
    \bigl\{ \, \efN[\W]  \,\mid\, \efN \in \Nmaps \, \bigr\}  \;\,=\;\,  
    \bigl\{ \, x \in \R  \,\mid\, \efNm[\W] \le \,x\, \le \, \mN[\W] \, \bigr\}  
    \label{eq:070}              
\end{equation}

\medskip
\noindent We wish to emphasize the following points regarding this Theorem.

\medskip

(a) Since $\Nmaps$ is non-empty, the set of all possible effective number assignments 
for a given counting vector $\W$ (LHS of \eqref{eq:070}) is also non-empty, 
and thus equal to closed interval 
$[\, \efNm[\W] \,,\, \mN[\W] \,]$.\footnote{As shown in \cite{Hor18A}, this closed interval 
shrinks to a single point $\{\, \efNm[\W]=\mN[\W] \,\}$ iff $\W$ is such that 
$\w_i \notin (0,1) \;,\, \forall\, i$.}

\smallskip

(b) It follows from \eqref{eq:070} that
\begin{equation}
      \efNm[\W]   \,\le\,  \efN[\W]  \,\le\,  \mN[\W]  \quad , \quad  
      \forall\, \efN \in \Nmaps \;\;,\;\,  \forall\, \W \in \,\setW 
      \label{eq:080}                              
\end{equation}
Thus, the concept of effective counting measure necessitates the existence of a non-trivial intrinsic 
{\em minimal amount} (count, total), specified by $\efNm$. The universal (independent of $\nrN$) 
function $\cfu$ entering the definition \eqref{eq:065} of this minimal ENF is referred to as 
the {\em minimal counting function}. This result has non-trivial consequences, including those 
concerning quantum uncertainty discussed here.

\smallskip

(c) In addition, the Theorem conveys that $\efNm$ is the only ENF with such definite structural 
role in $\Nmaps$. For example, since $\mN \notin \Nmaps$, there is no largest element in 
$\Nmaps$, i.e. the analog of $\efNm$ ``at the top". More importantly, for given $\W$, function 
$\efN$ can be adjusted to accommodate any intuitively possible total larger than $\efNm[\W]$. 
Consequently, there are no ``holes" in the bulk of $\Nmaps$ where other privileged ENFs could 
be identified. Given its absolute meaning, the minimal amount $\efNm$ provides for the canonical 
solution of the quantum identity problem~\eqref{eq:010}. In particular \cite{Hor18A}
\begin{equation}
    \ketpsi   \; , \;  \{ \, \keti \}  \qquad \longrightarrow \qquad    
    \efNm[\, \ketpsi ,\{ \keti \} ]  \;=\;   \efNm[\W]
    \qquad , \qquad \w_i = N \mid \!\langle i \mid \psi \rangle \!\mid^2  
    \label{eq:090}             
\end{equation}

%%%%%%%%%%%%%%%%%%%%%%%%%%%%%%%%%%%%%%%%%%
\section{The Measure Aspect of Quantum Uncertainty}

The uncertainty in QM refers to the indeterminacy of outcomes obtained by probing the quantum state. 
More specifically, consider a canonical situation where state $\ketpsi$ from $\nrN$-dimensional Hilbert 
space is probed by measuring the observable associated with non-degenerate Hermitian operator $\opO$. 
In a standard manner, $\ketpsi$ is repeatedly prepared and measured, generating a sequence of outcomes
\begin{equation}
    \mid \! \psi \,\rangle  \quad 
    \xrightarrow{\; \text{measure} \; \opO \;} \quad
    \{ \, (\, \mid \! i_\ell \,\rangle , O_{i_\ell} \,) \,\mid\, \ell =1,2,\ldots \, \}
    \label{eq:100}              
\end{equation}
With $\{\, (\, \mid \! i \,\rangle , O_i \,) \mid i=1,2,\ldots ,\nrN \}$ denoting the set of eigenstate-eigenvalue 
pairs, $(\, \mid \! i_\ell \,\rangle , O_{i_\ell}\,)$ specifies the outcome of $\ell$-th trial, namely the collapsed 
state and the measured value. Quantum uncertainty of $\ketpsi$ with respect to its probing by $\opO$ is 
intuitively associated with the ``spread" of outcomes $(\, \mid \! i_\ell \,\rangle , O_{i_\ell}\,)$ so generated.

Clearly, the precise content of the notion so construed depends on how we choose to quantify the ``spread". 
In the usual approach, the focus is on the sequence of eigenvalues $O_{i_\ell}$ with spread characterized 
in terms of distance (metric) on the spectrum of $\opO$. We will refer to this approach as 
{\em metric uncertainty} ($\rho$-uncertainty). A commonly used quantifier of this type is the standard 
deviation which leads to a particularly simple form of quantum uncertainty relations \cite{Hei27A}.
 
However, it may be interesting to view quantum indeterminacy differently~\cite{Hor18B}. 
A possible approach is to characterize it by the effective number of distinct outcomes occurring in 
$\eqref{eq:100}$. This would express the spread in terms of the ``amount", and we refer to it as
{\em measure uncertainty} ($\mu$-uncertainty). Note that, in this case, it is immaterial whether we focus 
on sequence $O_{i_\ell}$, sequence $\mid \! i_\ell \,\rangle$, or on the sequence of corresponding pairs: 
the object of interest is the effective total of the outcomes. While such approach might have seemed rather 
nebulous in the past, it is clear that the effective number theory not only puts it on a firm ground, but also 
leads to rather unexpected revelations.

First, by construction, the set $\Nmaps$ of ENFs is identical to the set of all possible $\mu$-uncertainties. 
More explicitly, if $\mu=\mu[\, \ketpsi , \{ \,\keti \} ]$ formally denotes a valid $\mu$-uncertainty map, 
then there exists $\efN \in \Nmaps$ such that for all $\ketpsi$ and $\{ \,\keti \}$
\begin{equation}
    \mu[\, \ketpsi, \{ \,\keti \} ]   \;=\;  \efN[\W]   \qquad , \qquad
    \W = (\w_1, \ldots, \w_\nrN)   \qquad , \qquad 
    \w_i = N \mid \!\langle i \mid \psi \rangle \!\mid^2  
     \label{eq:110}             
\end{equation}
and vice versa. In other words, $\mu=\mu[\, \ketpsi , \{ \,\keti \} ]$ is the same object as 
$\efN=\efN[\, \ketpsi ,\{ \,\keti \} ]$ featured in the quantum identity problem \eqref{eq:010}.

Secondly, the existence of minimal effective number gives rise to minimal $\mu$-uncertainty. 
In particular, the effective number theory allows us to deduce the following rigorous statement  
\begin{description}
     \item{[U$_0$]}  
     {\em $\,$ The $\mu$-uncertainty of $\ketpsi$ with respect to $\{ \, \keti \}$ is at least 
     $\efNm[\, \ketpsi ,\{ \,\keti \} ]$ states. $[\,$Eqs.~\eqref{eq:065}, \eqref{eq:090}$\,]$.}  
\end{description} 
Remarkably, U$_0$ asserts that uncertainty is built into quantum mechanics as an absolute concept. 
In particular, by expressing it as a measure, we learn that there exists an intrinsic irremovable 
``amount" of uncertainty in state $\ketpsi$ relative to probing basis $\{\, \keti \}$, specified uniquely 
as $\efNm[\, \ketpsi ,\{ \keti \} ]$ states.

Note that U$_0$ can be viewed as a quantum uncertainty principle of very different kind than 
the one conveyed by Heisenberg relations~\cite{Hei27A,Ken27A}. 
Indeed, while the latter is of a relative (comparative) nature, the $\mu$-uncertainty principle is 
absolute. It allows us to express the fundamental difference between quantum and classical 
notions of state in a particularly direct and economic way represented by the following diagram
\begin{displaymath}
   \begin{aligned}
     \text{classical state} \; S  
     \qquad\qquad\quad  &\leftrightarrow \qquad\qquad\quad                
     \text{quantum state} \; \ketpsi  \\[7pt]
     S \; \xrightarrow{\;\, \text{measurement} \;\,} \; S  
     \qquad\qquad &\leftrightarrow \qquad\qquad\;\,           
     \ketpsi  \; \xrightarrow{\;\, \text{measurement} \;\,} \;  \keti  \\[7pt]
      \mu\text{-uncertainty} \,=\,1 \; \text{state}
     \qquad\;\; &\leftrightarrow \qquad 
     \mu\text{-uncertainty} \,=\, \efNm[\, \ketpsi ,\{ \,\keti \} ]  \; \text{states}      
   \end{aligned}
\end{displaymath}
Thus, the properties of classical state $S$ can be measured with arbitrarily small error, meaning 
that its intrinsic $\mu$-uncertainty is always one (S has single identity). On the other hand, 
if the probing of a quantum state $\ketpsi$ involves the collapse into elements of basis 
$\{\, \keti \}$, its intrinsic $\mu$-uncertainty is $\efNm[\, \ketpsi ,\{ \,\keti \} ]$, which is generically 
much larger than one ($\,\ketpsi$ has $\efNm[\, \ketpsi ,\{ \,\keti \} ]$ identities).

Finally, it is important to point out that the above considerations are by no means restricted 
to \added[G]{a} discrete case or finite-dimensional Hilbert spaces. The relevant extensions are 
worked out in complete generality by Ref.~\cite{Hor18B}. As an elementary example, consider 
the case of a spinless Schr\"odinger particle contained in region $\sdm \subset \R^D$ with volume 
$V$. Using the above results, one can show that its minimal $\mu\,$-uncertainty with respect 
to the position basis is given by 
\begin{equation} 
   \efVs_\star[\psi] \,=\, 
   \int_\sdm \, d^D x \; \min \{ V \psi^\star(x)\psi(x)\, , 1 \}
   \label{eq:u120}                   
\end{equation}
where $\psi(x)$ is particle's wave function. Thus, the measure uncertainty takes the form of 
{\em effective volume}. The corresponding $\mu$-uncertainty principle states that a particle 
described by $\psi$ cannot be associated with an effective volume smaller than 
$\mathscr{V}_\star[\psi]$. 

%%%%%%%%%%%%%%%%%%%%%%%%%%%%%%%%%%%%%%%%%%
\section{Conclusions}

In the first part of this presentation, we outlined the construction of effective number theory 
and discussed how it solves the quantum identity problem~\cite{Hor18A}. The latter has 
a wide range of potential applications in quantum physics. Particularly close attention 
was given to the most consequential result of effective number theory, namely the existence 
of a minimal amount (total, count) consistently assignable to a collection of objects 
distinguished by probability weights. This finding offers a rather unexpected new 
insight into the nature of measure.

In the second part of the presentation, we analyzed the consequences of effective number 
theory for the concept of uncertainty in quantum mechanics~\cite{Hor18B}. In particular, 
we argued that the ensuing measure approach reveals the existence of uniquely-defined 
intrinsic $\mu$-uncertainties in quantum state, each associated with a particular way of 
probing it. We propose these intrinsic uncertainties as potentially useful characteristics 
of quantum states. It is interesting to note in this regard that, starting from essentially 
classical (measure) considerations, we arrived at describing aspects of the state 
that are truly quantum in their nature.

\bigskip
\noindent
{\bf Acknowledgments.} I.H. acknowledges the support by  the Department of Anesthesiology at
the University of Kentucky.

%%%%%%%%%%%%%%%%%%%%%%%%%%%%%%%%%%%%%%%%%%

%%%%%%%%%%%%%%%%%%%%%%%%%%%%%%%%%%%%%%%%%%
\end{document}